\pdfoutput=1

\documentclass[conference]{IEEEtran}
\ifCLASSINFOpdf
\else
\fi

\usepackage{xcolor}
\usepackage{cite}
\usepackage{amsmath,amssymb,amsfonts}
\usepackage{graphicx}
\usepackage{multirow}
\usepackage{soul}

\usepackage{cite}
\usepackage{amsmath,amssymb,amsfonts}
\usepackage{graphicx}
\usepackage{textcomp}
\usepackage{xcolor}
\usepackage{algorithm}
\usepackage{algpseudocode}
\usepackage{latexsym}
\usepackage{amsmath}
\usepackage{graphicx}

\begin{document}
%
\title{LogQA: Question Answering in Unstructured Logs}
%
%
%

	\author{
	    \IEEEauthorblockN{
	    Shaohan Huang\IEEEauthorrefmark{1}, 
	    Yi Liu\IEEEauthorrefmark{1},
	    Carol Fung\IEEEauthorrefmark{2},
	     Jiaxing Qi\IEEEauthorrefmark{1},
	     Hailong Yang\IEEEauthorrefmark{1},
	     Zhongzhi Luan\IEEEauthorrefmark{1}
	     }
	     \\
	     \IEEEauthorblockA{\IEEEauthorrefmark{1}Sino-German Joint Software Institute, Beihang University, Beijing, China
	    }
	    \\
	    \IEEEauthorblockA{\IEEEauthorrefmark{2}Computer Science Department, Virginia Commonwealth University, Richmond, Virginia, USA
	    }
	    \\
	    \IEEEauthorblockA{\{huangshaohan, yi.liu, luan.zhongzhi\}@buaa.edu.cn}
	}
\maketitle

\begin{abstract}

Modern systems produce a large volume of logs to record run-time status and events. System operators use these raw logs to track a system in order to obtain some useful information to diagnose system anomalies. One of the most important problems in this area is to help operators find the answers to log-based questions efficiently and user-friendly. In this work, we propose LogQA, which aims at answering log-based questions in the form of natural language based on large-scale unstructured log corpora. Our system presents the answer to a question directly instead of returning a list of relevant snippets, thus offering better user-friendliness and efficiency. 
LogQA represents the first approach to solve question answering in lod domain.
LogQA has two key components: \textit{Log Retriever} and \textit{Log Reader}. Log Retriever aims at retrieving relevant logs w.r.t. a given question, while Log Reader is responsible for inferring the final answer. Given the lack of a public dataset for log questing answering, we manually labelled a QA dataset of three open-source log corpus and will make them publicly available. We evaluated our proposed model on these datasets  by comparing its performance with 6 other baseline methods. Our experimental results demonstrate that LogQA has outperformed other baseline methods. 


\end{abstract}

\begin{IEEEkeywords}
Log data analysis, question answering, information retrieval
\end{IEEEkeywords}

%
\IEEEpeerreviewmaketitle

\section{Introduction}
%
%
%
%
\IEEEPARstart{M}{odern} Computer systems have become increasingly complex as service systems grow in both size and complexity~\cite{tan2010predictability}. These systems produce a large volume of logs, which are widely used to record run-time status and events.  Since system logs contain noteworthy information, they have become one of the most important data sources for system monitoring to improve service quality.

\begin{figure}[t]
	\begin{center}
		\includegraphics[width=7.cm]{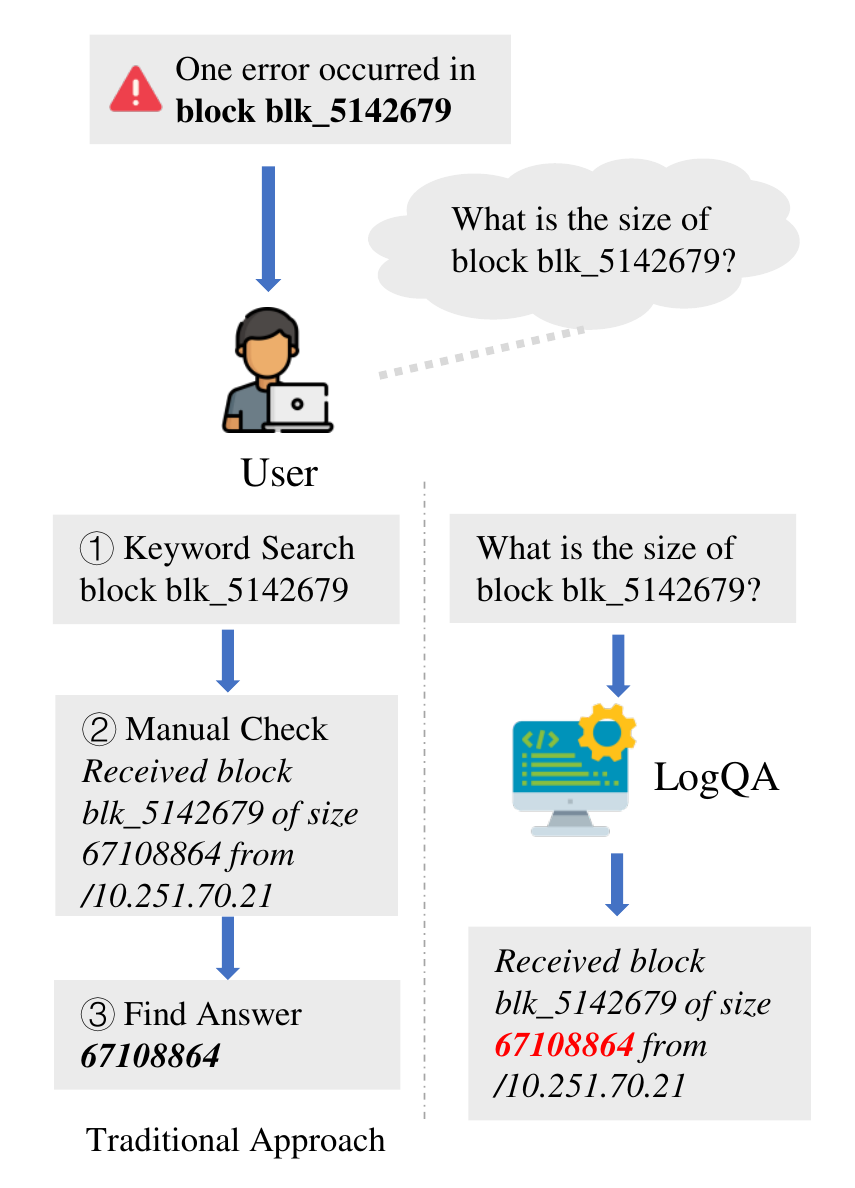}
	\end{center}
	\caption{Compared to traditional approach, LogQA directly answers log-based questions in the form of natural language.}
	\label{fig:demo}
\end{figure}

Many log analysis approaches have been proposed for service management. As summarized in~\cite{he2021survey}, automated log analysis can be  classified into log compression (e.g.,~\cite{burrows1992line}, ~\cite{skibinski2007fast}), log parsing (e.g,~\cite{zhu2019tools}, \cite{he2017drain}, \cite{huang2020paddy}), anomaly detection (e.g,~\cite{zhang2019robust}, \cite{huang2020hitanomaly}), failure 
diagnosis~(e.g.,~\cite{tanwir2015information,kim2016fault}), and failure prediction~\cite{zheng2009system,zhang2016automated}. The existing studies have boosted the effectiveness and efficiency of systematic usage of service logs. However, operators often have to check the raw logs in order to acquire some information or diagnose system events. 
Figure~\ref{fig:demo} shows an example, an HDFS (a java based distributed file system used in Hadoop) system administrator discovered an error in block blk\_5142679. He wants to figure out the size of this block. In the traditional processes, he begins by utilizing the \textit{block blk\_5142679} to search for relevant system logs in the entire logs. Then, he carefully checks each log entry to identify the answer. One of the most critical challenges in this area is assisting operators in locating answers in an efficient and user-friendly way. However, manual or the rule-based log question answering has become ineffective and inefficient. First, modern computer systems produce a large volume of logs (e.g., some cloud systems would produce about 30-50 gigabytes of tracing logs per hour~\cite{zhu2019tools}). Locating an answer manually becomes time-consuming. Secondly, developers may frequently modify the source code of log recording, which leads to log data format change and the occurrence of some unseen log events, which makes it difficult to handle with rule-based methods as well.
Thirdly, many services are implemented and maintained by a number of operators. Operator with domain knowledge is expected to find answers through log examination. How to construct a user-friendly and effective solution to assist system administrators in obtaining important information and locating the answers to log-based questions has not yet been thoroughly investigated.


Question answering is an important topic in the field of natural language processing, and many works have been proposed recently~\cite{devlin2018bert}.
However, there are some challenges in using NLP algorithms for log question answering. First, there exists a domain shift between general natural language and log data. Log data includes domain-specific symbols, such as IP addresses and modular identifiers~\cite{he2017towards}. General NLP techniques consider these symbols to be out-of-domain terms and replace them with a special token, however they are crucial for the log domain and cannot be ignored.
Second, applying NLP methods to learn a question answering model usually requires a large-scale training set~\cite{devlin2018bert}. However, to the best of our knowledge, there is no public question answering dataset. It is difficult to implement an effective question answering system with limited data. 

In this work, we propose a question answering system for unstructured logs, namely LogQA, which aims to answer questions in the form of natural language based on large-scale unstructured log corpus. LogQA has two key components: Log Retriever and Log Reader. Log Retriever retrieves some relevant and helpful raw logs. The goal of Log Reader is to predict exact answers based on retrieved logs. We follow~\cite{goldberg2014word2vec} to involve negative sampling to train Log Retriever. Due to the limitation of our training data size, there exists many completely unrelated logs in negative examples, which cannot provide sufficient informative examples for training. To address the above challenge, we propose an iterative method to generate hard negative examples into our training process.

Due to the lack of publicly available question answering benchmarks, we manually labeled a QA dataset over three public log datasets (HDFS, OpenSSH, and Spark) and will make them public available. We evaluate our proposed method on these datasets. Experimental results show that Log Retriever outperforms other existing log-based retrieval methods. The accuracy of our top-5 model is over 30 points higher than the one ranked second in an extremely difficult dataset. We also investigate the impacts of hard negative in training and the number of training iteration. In log reading task, we compare the performance of Log Reader with four baseline methods. Experimental results show that our method achieves the best performance on three datasets.

The key contributions of this paper can be summarized as follows:
\begin{itemize}
\item  We propose LogQA, a framework to answer a question in the form of natural language based on large-scale unstructured log corpus, which is easy to help operators find the answer efficiently and user-friendly.
\item LogQA consists of log retrieving and log reading. Both achieve the state-of-the-art results.
\item We design a hard negative sampling into Log Retriever training process, which is able to improve the robustness and efficiency of retrieval model.
\item We manually annotate a QA dataset for three open-source log datasets and will make them public available, which will benefit the research community  in log-related QoS area.
\end{itemize}

The rest of the paper is organized as follows. We introduce the overview of our work in Section~\ref{sec:overview}. The detail of our model are described in Section~\ref{sec:method}. Section~\ref{sec:exp} describes the experimental settings, and log question answering dataset we manually labelled. We evaluate the performance of our method~\ref{sec:results}. Related work is introduced in Section~\ref{sec:related}. Finally, in Section~\ref{conc}, we conclude our work and state possible future work.

\begin{figure}[t]
	\begin{center}
		\includegraphics[width=7.5cm]{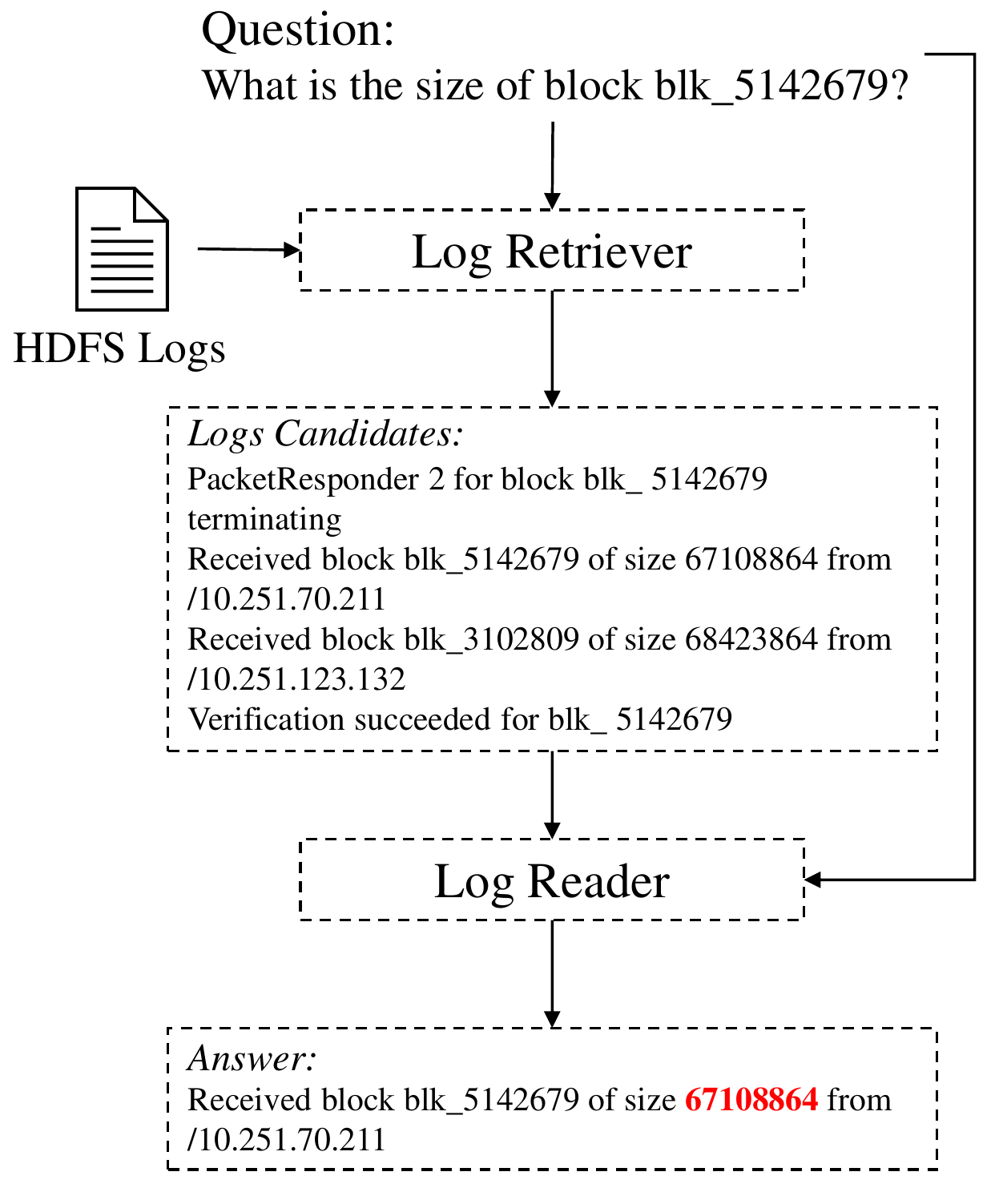}
	\end{center}
	\caption{Workflow of LogQA system.}
	\label{fig:overview}
\end{figure}

\section{Overview}\label{sec:overview}
LogQA takes a question $x$ as input and learns a distribution $p(y|x)$ over outputs answer $y$. We decompose $p(y|x)$ into two steps: \textit{retrieve}, then \textit{read}. Given a question $x$, we first retrieve some possible relevant logs $z$ from a raw log corpus $\mathcal{Z}$. We call this step as a distribution $p_1(z|x)$. Then, based on retrieved logs $z$ and question $x$ to output the answer $y$. This step is modeled as $p_2(y|z,x)$. The overall likelihood can be defined as:
\begin{equation}
    p(y|x) = \sum_{z \in \mathcal{Z}}{p_2(y|z,x)p_1(z|x)}
\end{equation}

Therefore, LogQA has two key components: Log Retriever, which models $p_1(z|x)$ and Log Reader, which models $p_2(y|z,x)$. The overall framework is illustrated in Figure~\ref{fig:overview}. For example, an HDFS IT operator asks a question \textit{`What is the size of block blk\_5142679?'}. Log Retriever retrieves some relevant and helpful raw logs from an HDFS log corpus (e.g., \textit{Received block blk\_5142679 of size 67108864 from /10.251.70.211}). We regard these retrieved logs as log candidates and feed them into Log Reader. The goal of Log Reader is to predict exact answers for users. In this case, Log Reader extract the correct answer \textit{67108864} from the log \textit{Received block blk\_5142679 of size 67108864 from /10.251.70.211}. 

In the training phase, the Log Retrieval and Log Reader models are optimized, respectively. During the inference phase, we pass the question through Log Retrieval to retrieve the log candidates, then utilize the question and the log candidates to determine the exact answer.

\section{Methodology}\label{sec:method}
In this section, we will first describe the log parsing employed in this research. Then, we describe our Log Retriever's design. Then, we present our Log Reader model in detail.

\subsection{Log Parsing}\label{sec:logparsing}
Log parsing is a crucial tool for processing unstructured, free-text logs for many log-based applications such as anomaly detection (e.g,~\cite{zhang2019robust}) and failure diagnosis~(e.g.,~\cite{tanwir2015information,kim2016fault}). The process of parsing logs is also a very important component of LogQA.
The purpose of log parsers is to transform unstructured messages into structured log templates carrying key parameters. As shown in Table~\ref{table:logpariing}, one log message ``Received block blk\_-2856928563366064757 of size 67108864 from /10.251.42.9'' comes from  the HDFS dataset. It is parsed into log template ``Received block $\langle$*$\rangle$ of size $\langle$*$\rangle$ from /$\langle$*$\rangle$'' and parameter values [`blk\_-2856928563366064757', `67108864', `10.251.42.9']. The log template consists of fixed text characters, while parameters record system variables and properties.

\begin{table}[ht]
\renewcommand{\arraystretch}{1.8}
\setlength{\tabcolsep}{14pt}

    \caption{Example of log parsing.
    }
    \centering
    \label{table:logpariing}
    \centering
    \begin{tabular}{p{0.15\linewidth}  p{0.6\linewidth}}
    \hline
   Raw log       &  Received block blk\_-2856928563366064757 of size 67108864 from /10.251.42.9 \\ \hline 
   Template   &	 Received block $\langle$*$\rangle$ of size $\langle$*$\rangle$ from /$\langle$*$\rangle$ \\ \hline
   Parameters &	`blk\_-2856928563366064757', `67108864', `10.251.42.9'   \\ \hline

    \end{tabular}
\end{table}

There have been many studies on log parsing~\cite{yamanishi2005dynamic,oliner2012advances,niwattanakul2013using,he2016evaluation,he2017drain}. Therefore, log parsing is not the focus of this paper. Considering accuracy and speed, we adopt Drain~\cite{he2017drain} method to parse all log data.

\subsection{Log Retriever}\label{sec:logretiever}
In the Log Retriever part, we first introduce the architecture model of Log Retriever. Then we describe the training method of Log Retriever. 

\subsubsection{Model Architecture}
Given a question $x$, the goal of Log Retriever is to retrieve potentially helpful logs $z$ from a large log corpus. The retrieved logs contain a span that can answer the question. The Log Retriever can be defined as:
\begin{equation}\label{equa:logretrieve} \begin{split}
    p(z|x) & = \frac{exp f(x, z)}{\sum_{z_i}{exp f(x, z_i)}} \\
    f(x, z) & = cosine(encode(x), encode(z))
\end{split} \end{equation}
where $f(x, z)$ is a function to compute the similarity between $x$ and $z$. We use an encoder to convert a question or log into a vector, then use cosine distance to measure the similarity between $x$ and $z$.  
Cosine similarity is a measure of similarity between two non-zero vectors defined in an inner product space. It is used as a distance evaluation metric between two points in the plane. It is a widely used metric in machine learning algorithms like the $k$-NN for determining the distance between the neighbors in information retrieval~\cite{he2021survey}.
The retrieval distribution is the softmax over all similarity scores.

\begin{figure}[t]
	\begin{center}
		\includegraphics[width=8.cm]{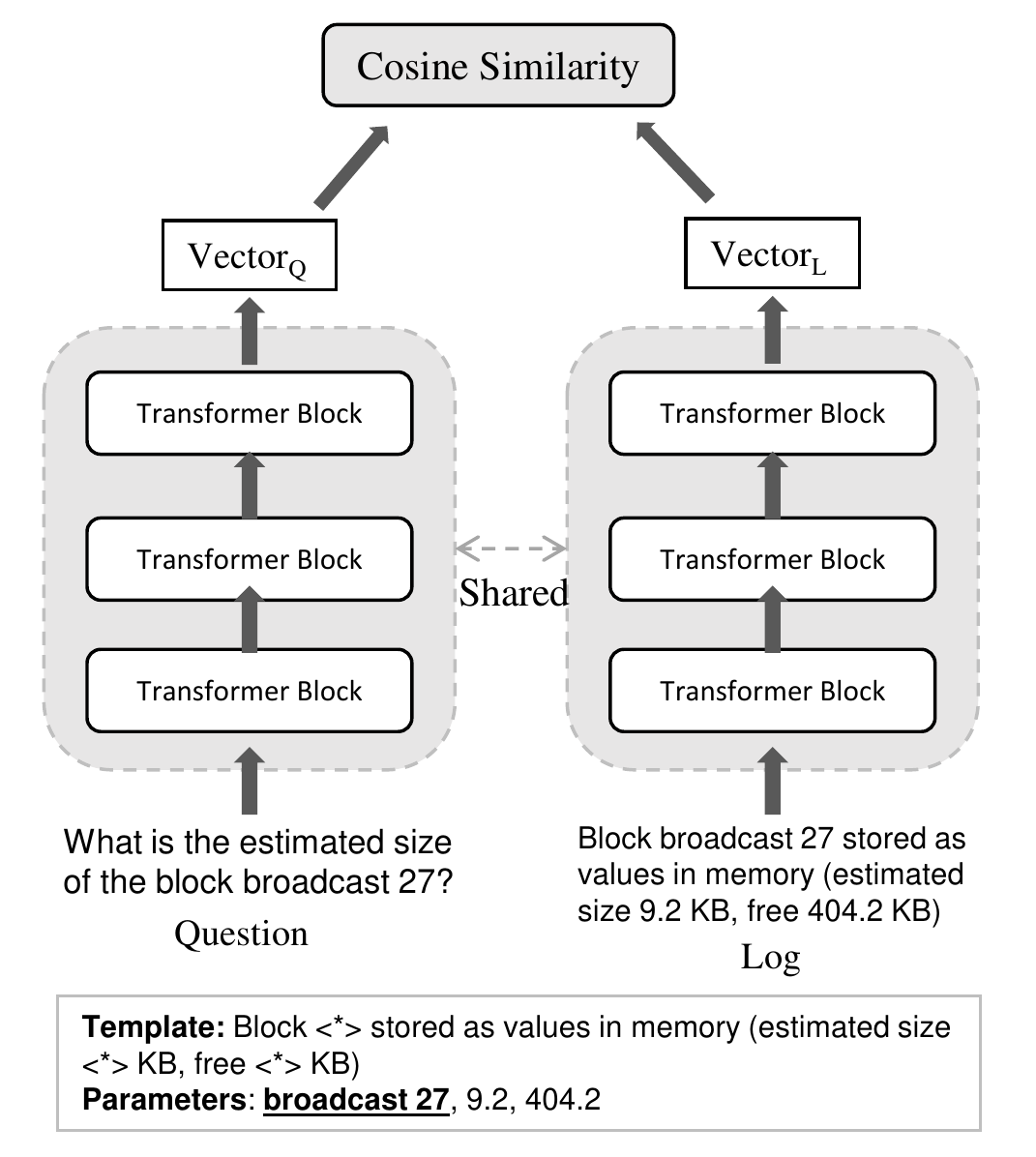}
	\end{center}
	\caption{Log Retriever Model.}
	\label{fig:model}
\end{figure}

The similarity function $f(x, z)$ is the core structure of our Log Retriever. As shown in Figure~\ref{fig:model}, we design a two-tower structure, where each tower is a multi-depth Transformer-based~\cite{vaswani2017attention} encoder to learn the representations of questions and raw logs. Given a question $x$ and a log $z$, the left encoder is to learn the contextual representations of question $x$ and the right encoder converts the raw log $z$ into a dense vector. Then, we use cosine function to compute the similarity between the question $x$ and the raw log $z$. Because we only have limited labeled data as our training data, we share the parameters of encoders in these two towers.

\subsubsection{Retriever Training}
Now we present the training process of Log Retriever. As shown in the Equation~\ref{equa:logretrieve}, we train Log Retriever model by maximizing the likelihood $p(z|x)$. The key computational challenge is that the probability $p(z|x) = \frac{exp f(x, z)}{\sum_{z_i}{exp f(x, z_i)}}$, where $z_i \in \mathcal{Z}$, involves a summation over all log corpus $\mathcal{Z}$. The log $z$ that contains the correct response is considered potentially helpful. In addition, we discover that some logs may not contain the correct answer, but some of their parameters exist in question. In the log retriever, these question-related parameters also contain crucial information. Therefore, we customize the following three target functions:
\[
\mathbf{T}(z) = \left\{%
 \begin{array}{l}
  1 \text{  if the log $z$ contains answer y;}\\
  \alpha \text{  if the log $z$ contains parameters of $x$;}\\
  0 \text{  otherwise.}
 \end{array}%
\right.
\]
where $\mathbf{T}(z)$ denotes the target value of the probability $p(z|x)$. In our work, we set the $\alpha$ value as 0.2.

Inspired by~\cite{goldberg2014word2vec}, we involve negative sampling into our training. More specifically, we sample a set of logs from raw log corpus $\mathcal{Z}$ and use these sampled logs as negative examples. The shortcoming of the negative sampling method is that these logs are randomly sampled from the entire corpus and there exist some completely unrelated logs that will be used as negative examples. However, these instances are too simple for the similarity model. To overcome this problem, we propose an iterative method to generate hard negative examples into our training process.

The training algorithm is explained in Algorithm~\ref{alg:cap}. We first initialize our Log Retriever model with BERT~\cite{devlin2018bert}, which is a pre-trained bidirectional transformer language model. The BERT model can help the Log Retriever model capture more semantic information and better understand questions in the form of natural language. After initialization, we sample question log pairs from labeled data $\mathcal{P}$ and use these data to optimize our model. After finishing each iteration, we use our model to obtain hard negative examples. Given questions in labeled data, we use our model to retrieve log candidates from raw log corpus and then remove the logs which contain the correct answer from log candidates. These remaining candidates cannot answer the question but are very similar to the question. We label these candidates as hard negative samples and add them to our training data. Therefore, we extend our training objective defined in Equation~\ref{equa:logretrieve} to incorporate hard negative examples as follows:
\begin{equation}\label{equa:logretrieve_hard} \begin{split}
    p(z|x) & = \frac{exp f(x, z)}{\sum_{z_i \in \mathcal{B}}{exp f(x, z_i)} + w \sum_{z_j \in \mathcal{H}}{exp f(x, z_j)}} \\
\end{split} \end{equation}
where $\mathcal{B}$ represents a batch of question log pairs and $\mathcal{H}$ is  hard negatives. $w$ is the weight of hard negative examples. We train Log Retriever with different values $w$ and compare the impact of this weight in our experiments.

There are two advantages to our proposed method. First, our negative examples are more informative and reduce the impact of extremely easy negative examples. Second, our hard negative sampling can be regarded as a kind of data augmentation method, which can solve the insufficient training data problem.

\begin{algorithm}
\caption{Log Retriever Training}\label{alg:cap}
\begin{algorithmic}
\Require Question Log Pairs $\mathcal{P}$, Raw Log Corpus $\mathcal{L}$, Max Iteration Number $I$
\State Initialize log retriever model $M$ with BERT
\State $i \gets 0$
\While{$i < I$}
    \For{Batch $\mathcal{B}$ in $\mathcal{P}$}
    \State Compute loss based Equation~\ref{equa:logretrieve}
    \State Update model M
    \EndFor
    \State \textit{//obtain hard negative examples}
    \For{ each pair $(q, l)$ in $\mathcal{P}$} 
    \State Retrieved Logs $L = M(q)$
    \State Remove $l$ from $L$
    \State $\mathcal{P} \gets \mathcal{P} + L$
    \EndFor
    \State $i \gets i + 1$
\EndWhile
\State \textbf{Return} log retriever model $M$
\end{algorithmic}
\end{algorithm}

\subsection{Log Reader}\label{sec:logreader}
In this section, we introduce the design of Log Reader, which is the other core component of the LogQA system. It aims at inferring the exact answer in response to the question from a set of retrieved logs. 

\begin{figure}[t]
	\begin{center}
		\includegraphics[width=8.5cm]{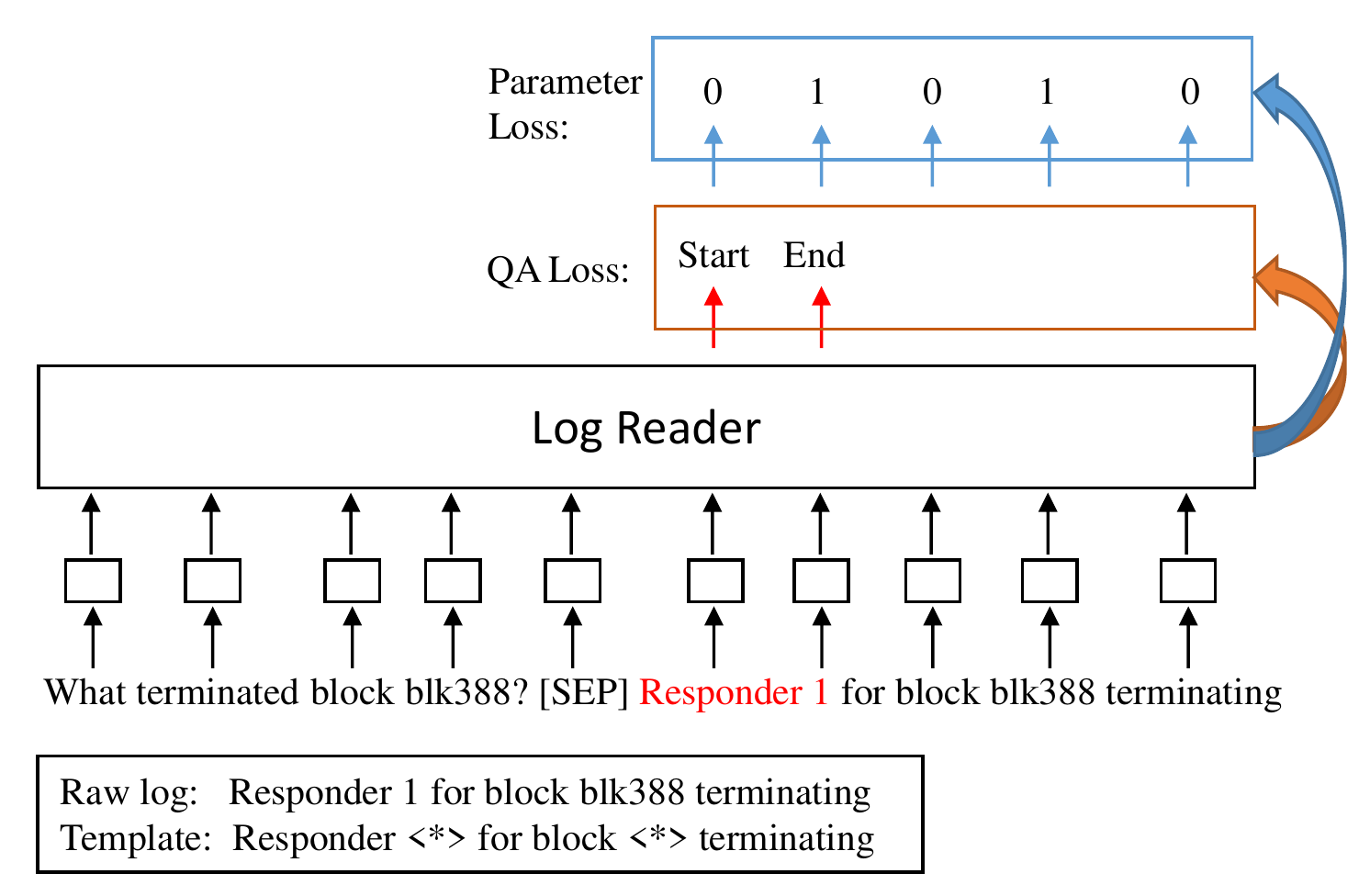}
	\end{center}
	\caption{Log Reader Model.}
	\label{fig:reader_model}
\end{figure}

Given the top $k$ retrieved logs, the Log Retriever assigns a selection score to each retrieved log. In addition, Log Reader extracts an answer span from each log and assigns a span score. In labeled training data, the correct answer to a given question must exist in the log. Following~\cite{devlin2018bert}, the Log Retriever model focuses on learning how to predict the start and end position of an answer span from the retrieved logs. As shown in Figure~\ref{fig:reader_model}, we concatenate the input question (\textit{What terminated block blk388?}) and the context log (\textit{Responder 1 for block blk388 terminating}) as a single packed sequence and feed the packed sequence into the Log Reader model. In the answer \textit{Responder 1}, \textit{Responder} is the starting position and \textit{1} is the ending token.
Specifically, let $h_i$ be a word representation for $i$-th token in the retrieved log. The probabilities of a token being the
starting/ending positions of an answer span are defined as:
\begin{equation}\label{equa:logreader} \begin{split}
    P_{start,i} & = softmax(h_i \cdot W_{start}) \\
    P_{end,i} & = softmax(h_i \cdot W_{end}) 
\end{split} \end{equation}
where $W_{start}$ and $W_{end}$ are learnable parameters. We compute a span score of the $s$-th to $e$-th words from the retrieved log as $P_{start,s} \times P_{end,e}$. 

We design two training objective functions to optimize our Log Reader as following:
\begin{equation}
    \mathcal{L} = \mathcal{L}_{QA} + \mathcal{L}_{param}
\end{equation}
The training objective $\mathcal{L}_{QA}$ is to maximize the marginal log-likelihood of all correct answer spans in the retrieved logs. The objective $\mathcal{L}_{param}$ is to predict whether the token represents a parameter variable. As shown in Figure~\ref{fig:reader_model}, the raw log is \textit{Responder 1 for block blk388 terminating} and its parameters are \textit{1} and \textit{blk388}. The goal of $\mathcal{L}_{param}$ is to find parameters \textit{1} and \textit{blk388} in raw log.
The training objective $\mathcal{L}_{param}$ has two advantages. First, parameters are typically used to record system variables and properties, and they are more likely to contain the answer to the question. Second, due to the limited data, only deploying QA loss cannot efficiently optimize Log Reader model. The parameter loss can help model understand log data better.

In our inference stage, we combine the probability of a log containing the answer and that of a token being the starting and ending position of an answer span, and selects the answer with the highest probability. 

\section{Experiment Setting}\label{sec:exp}
In this section, we present our experimental setting and evaluation results of LogQA. We first describe how to build unstructured logs question answering datasets. Then we introduce our evaluation metrics to evaluate QA systems. Finally, we describe some baseline methods and the hyperparameters in our experiments.

\subsection{Datasets}
We conduct question answering experiments on three public log datasets: HDFS dataset~\cite{xu2009detecting}, OpenSSH dataset~\cite{huang2020paddy} and Spark~\cite{zhu2019tools}. Due to the lack of a publicly available question answering benchmark, we manually labeled the above public log datasets and make them publicly available~\footnote{https://github.com/LogQA-dataset/LogQA}. We select 2,000 logs per dataset as our raw log set. Then we use a question generation model\footnote{https://huggingface.co/iarfmoose/t5-base-question-generator} to generate reading comprehension-style questions~\cite{rajpurkar2016squad} with answers extracted from a log. For example, one raw log is \textit{Received block blk\_5142679 of size 67108864 from /10.251.70.211
}. We take \textit{67108864} as its answer and feed it into the question generation model. The model will output \textit{What is the size of block blk\_5142679?}. Since we don't know which span is a proper answer, we iterate through all tokens in a log as an answer to generate questions. We collected more than 10,000 questions per dataset and then labeled all questions manually. The detailed information of the above datasets is listed in Table~\ref{table:data}. We can find that we keep only 2-3\% of the data after human annotation. A possible reason is that the question generation model we used is trained on Wikipedia-based question answering data~\cite{rajpurkar2016squad} and there exists a domain shift between Wikipedia and log data.

\begin{table}[ht]
\renewcommand{\arraystretch}{1.8}
\setlength{\tabcolsep}{14pt}

    \caption{Statistics of log QA data.
    }
    \centering
    \label{table:data}
    \centering
    \begin{tabular}{ccc}
    \hline
   Datasets       & \# of QA pairs    &  Length of Question \\ \hline \hline
   HDFS   &	 247 &	7.83  \\ \hline
   OpenSSH &	188  &	9.03  \\ \hline
   Spark &	397  &	7.00  \\ \hline

    \end{tabular}
\end{table}

To present the diversity of unstructured logs QA data, we categorize these questions into different types. As shown in Table~\ref{table:type}, in HDFS and OpenSSH datasets, the most common question type is `what', which accounts for roughly 90\% of all questions. The Spark data is more diverse and contains six different types of questions. In the Spark data, the most common question type is `how many'.

\begin{table}[ht]
\renewcommand{\arraystretch}{1.8}
    \caption{Question Type of log QA data.
    }
    \centering
    \label{table:type}
    \centering
\begin{tabular}{|lr|lr|lr|}
\hline
\multicolumn{2}{|c|}{HDFS} & \multicolumn{2}{c|}{OpenSSH} & \multicolumn{2}{c|}{Spark} \\ \hline \hline
What         & 88.3\%      & What           & 92.0\%      & How many      & 48.6\%     \\
Where        & 11.3\%      & Did            & 5.9\%       & What          & 27.0\%     \\
Others       & 0.4\%       & Who            & 1.1\%       & Is            & 18.9\%     \\
             &             & How many       & 1.1\%       & How large     & 2.8\%      \\
             &             &                &             & How Long      & 2.0\%      \\
             &             &                &             & Others        & 0.8\%     \\\hline
\end{tabular}
\end{table}


\subsection{Evaluation Metrics}
To measure the effectiveness of Log Retriever and Log Reader in question answering task, we use a set evaluation metrics to evaluate these two modules.

For the log retrieval part, we use \textit{top-k retrieval accuracy} (Acc@K), which are the most commonly used metrics in information retrieval problem~\cite{harman2011information}. The formal definition of Acc@K can be defined as follows: a log retriever $\mathcal{R}(q, \mathcal{C}) \xrightarrow{} \hat{\mathcal{C}}$ takes as input question $q$ and a log corpus $\mathcal{C}$ and returns a much smaller set $\hat{\mathcal{C}}$, where $\hat{\mathcal{C}} \in \mathcal{C}$ and $|\hat{\mathcal{C}}| = k 	\ll |\mathcal{C}|$. Top-k retrieval accuracy is the fraction of questions for which $\hat{\mathcal{C}}$ contains a span that can answer the question. In our experiments, we separately present the results of log retrieval where the $k$ is 1, 5 or 20.

For the log reading part, we model it into a reading comprehension task and follow~\cite{rajpurkar2016squad} and use \textit{Extra Match} (EM) and \textit{F1 score} (F1) as our evaluation metrics. EM measures the percentage of predictions that match the ground truth answer exactly. F1 score measures the average overlap between predictions and the ground truth answer. We consider the prediction and ground truth bags of tokens and compute their overlap F1 score. We compute these metrics as follows:
\begin{equation}
    \texttt{Precision} = \frac{\#Same}{\#Prediction}
\end{equation}
Precision shows the percentage of the same tokens among all text span predicted by the model.
\begin{equation}
    \texttt{Recall} = \frac{\#Same}{\#Ground truth}
\end{equation}
Recall means the percentage of the same tokens in the ground truth answer.
\begin{equation}
    \texttt{F1-Score} = \frac{2 \cdot \texttt{Precision} \cdot \texttt{Recall}}{\texttt{Precision} + \texttt{Recall}}
\end{equation}

For example, as shown in Figure~\ref{fig:overview}, the ground truth answer is `67108864'. We assume that the prediction of our model is `size 67108864'. The length of the same tokens `67108864' is 1 and we can compute that the precision is $\frac{1}{2}$ and recall is $1$. Its F1-score is $\frac{2}{3}$.

\begin{table*}[t]
\renewcommand{\arraystretch}{1.8}
\setlength{\tabcolsep}{12pt}
\centering
\caption{Evaluation results of log retrieval.
    }
    \centering
    \label{table:logre}
\begin{tabular}{|l|lll|lll|lll|}
\hline
                       & \multicolumn{3}{|c|}{\textbf{HDFS}}                 & \multicolumn{3}{c|}{\textbf{OpenSSH}}              & \multicolumn{3}{c|}{\textbf{Spark}}                \\ \hline \hline
                       & \textbf{Acc@1} & \textbf{Acc@5} & \textbf{Acc@20} & \textbf{Acc@1} & \textbf{Acc@5} & \textbf{Acc@20} & \textbf{Acc@1} & \textbf{Acc@5} & \textbf{Acc@20} \\ \hline
\textbf{Random}        & 0.0878         & 0.3243         & 0.6284          & 0.0625         & 0.2321         & 0.4464          & 0.0084         & 0.0672         & 0.1975          \\
\textbf{Edit Distance} & 0.4662         & 0.4865         & 0.5068          & 0.3304         & 0.3393         & 0.3571          & 0.0000         & 0.1387         & 0.2227          \\
\textbf{Jaccard}       & 0.2297         & 0.3176         & 0.3919          & 0.2411         & 0.3482         & 0.4554          & 0.2311         & 0.3193         & 0.4496          \\
\textbf{BM25}          & 0.2568         & 0.2635         & 0.2838          & 0.0179         & 0.2232         & 0.4375          & 0.0000         & 0.0210         & 0.0252          \\
\textbf{Jaro Winkler}  & 0.4730         & 0.5473         & 0.6689          & 0.2232         & 0.2589         & 0.3482          & 0.1933         & 0.1975         & 0.2437          \\
\textbf{BERT Cosine}   & 0.3649         & 0.6081         & 0.6486          & 0.3750         & 0.4464         & 0.5000          & 0.0798         & 0.1387         & 0.2437         \\ 
\textbf{Log Retriever}   &\textbf{0.5850}         & \textbf{0.7200}         & \textbf{0.7600}          & \textbf{0.5172}         & \textbf{0.6206}         & \textbf{0.6379}          & \textbf{0.3830}         & \textbf{0.6166}           & \textbf{0.8500}         \\ \hline
\end{tabular}
\end{table*}

\subsection{Baseline Methods}
We first introduce the baseline methods in the log retrieval part. These methods are briefly described as follows:
\begin{itemize}
\item Random: we randomly sample $k$ logs from raw corpus as our log retrieval results.
\item Edit Distance: we use edit distance to measure the similarity between question and raw logs.
\item Jaccard~\cite{niwattanakul2013using}: we use Jaccard similarity to measure similarities between questions and raw logs. Jaccard similarity can be computed as follows:
\begin{equation}
    J(Q, L) = \frac{|Q \cap L|}{|Q \cup L|}
\end{equation}
where $Q$ is a bag of question words and $L$ is the word set of a raw log.
\item BM25~\cite{robertson2009probabilistic}: BM25 is a ranking algorithm used by search engines to estimate the relevance of documents to a given search query, which is widely used in many information retrieval systems. We use Gensim\footnote{https://github.com/RaRe-Technologies/gensim} in our experiments.
\item Jaro Winkler~\cite{dressler2017efficient}: the Jaro–Winkler distance is various of Jaro distance. Jaro similarity is a string metric to measure an edit distance between two sequences. The Jaro–Winkler distance uses a prefix scale to give more favorable ratings to strings.
\item BERT Cosine~\cite{devlin2018bert}: BERT is a pre-trained model, which can be used to encode natural language text into dense vectors. We use BERT to encode question and raw logs respectively and use cosine function to measure their similarity.
\end{itemize}

We implement six baseline methods in the log reading task. These methods are described as follows:
\begin{itemize}
\item Random: we randomly sample one word from the retrieved log as its answer.
\item Slide Window: For each candidate answer, we compute the unigram overlap between the candidate answer and the question. Among these, we select the best one using the sliding-window approach proposed in~\cite{richardson2013mctest}
\item Logistic Regression~\cite{wright1995logistic}: We extract features for each candidate answer and use a logistic regression model as a binary classification model. To obtain a better representation of question and answer, we leverage the Log Retriever model to encode them and use hidden states as their feature.
\item BiDAF~\cite{seo2016bidirectional}: BiDAF is a machine comprehension model. It introduced a Bi-Directional Attention Flow (BIDAF) network, a multi-stage hierarchical process that represents the context at different levels of granularity.
\item QANet~\cite{yu2018qanet}: QANet consists exclusively of convolution and self-attention, where convolution models local interactions and self-attention models global interactions. 
\item BERT~\cite{devlin2018bert}: BERT is a pre-trained transformer-based model. It achieved many cutting-edge results in a variety of general question answering tasks in the field of natural language processing.
\end{itemize}

We implement Log Retriever and Log Reader based on PyTorch~\cite{paszke2017automatic} on the Linux server with NVIDIA Tesla V100 GPU. In our Log Retriever experiment, we set the weight of the hard negatives as 2 and the iteration number as 4. We set the learning rate as 5e-5. In our Log Reader experiment, we set the epoch as 15 and the learning rate as 3e-5. We randomly sample 60\% dataset as our training data, 10\% data for validation and others as testing data.

\section{Results}\label{sec:results}
Our experiments contain two sub-tasks: log retrieval and log reading. We first describe the experiment results of Log Retriever. We then show the performance of Log Reader on different log data. 
Lastly, we present the case study to help understand the difference between our model and other baseline methods.

\subsection{Log Retriever}
In this part, we first study the performance of Log Retriever on the three log datasets. Then we present the experiments of hard negative sampling. Lastly, we investigate the impacts of hard negative example weight and the number of training iterations.

\begin{table*}[t]
\renewcommand{\arraystretch}{1.8}
\setlength{\tabcolsep}{11pt}
\centering
\caption{Evaluation results of log retrieval without hard negative in training.
    }
    \centering
    \label{table:hardnega}
\begin{tabular}{|l|lll|lll|lll|}
\hline
                       & \multicolumn{3}{|c|}{\textbf{HDFS}}                 & \multicolumn{3}{c|}{\textbf{OpenSSH}}              & \multicolumn{3}{c|}{\textbf{Spark}}                \\ \hline \hline
                       & \textbf{Acc@1} & \textbf{Acc@5} & \textbf{Acc@20} & \textbf{Acc@1} & \textbf{Acc@5} & \textbf{Acc@20} & \textbf{Acc@1} & \textbf{Acc@5} & \textbf{Acc@20} \\ \hline
\textbf{Log Retriever}   & 0.5850        & 0.7200         & 0.7600          & 0.5172         & 0.6206         & 0.6379          & 0.3830         & 0.6166           & 0.8500         \\ 
\textbf{W/O Hard Negative}   & 0.3066        & 0.4266         & 0.5600          & 0.2730         & 0.3448         & 0.3965          & 0.3166         & 0.4583           & 0.7666         \\ \hline
\end{tabular}
\end{table*}

\subsubsection{Performance of Log Retriever} Table~\ref{table:logre} shows the performance of Log Retriever compared to six baseline methods over three log datasets. In the log retrieval task, Log Retriever achieves the highest accuracy among top-$1$, top-$5$, and top-$20$ over all datasets. On the HDFS dataset, the accuracy of random guesses is very low in the top $1$ (less than 10\%), which presents the quality and difficulty of this log retrieval task. We can find that our method has an accuracy score of more than 0.5 in the top 1 and of more than 0.7 in the top 5 and top 20. Compared to Edit Distance and Jaro Winkler methods, which don't take the semantic information into consideration, our method shows a higher accuracy. The results demonstrate that semantic information is crucial to log retrieval. For example, one question is \textit{``is the status of block blk\_5142679?''} and the correct log is \textit{``Verification succeeded for blk\_5142679''}. The word \textit{``succeeded''} can present the status of this block. But there is only one identical word between the question and the ground truth log, which may cause that similarity computed by Edit Distance or Jaro Winkler is small. We also observe that BERT Cosine performs worse than Edit Distance or Jaro Winkler. A possible explanation for this might be that BERT is trained in general natural language corpus and suffers from domain shift. 

A very similar conclusion emerges from the OpenSSH dataset. Log Retriever has an accuracy score of more than 0.5 in the top 1 and achieves the highest accuracy at top-$5$ and top-$20$. However, we observe that log retrieval is far more difficult in the Spark dataset. The accuracy of random guesses is less than 1\% in the top $1$. Edit Distance and BM25 cannot retrieve any correct log in the top $1$. A possible explanation for this might be that the Spark data is more diverse and contains more types of questions as shown in Table~\ref{table:type}. Our method achieves an accuracy score of 0.38 in the top $1$. Interestingly, we also observe that Log Retriever has achieved a good accuracy (0.85) at top-$20$. A possible explanation for these results may be that Log Retriever has high recall, but the recall rate is low facing various questions.

\subsubsection{Hard Negative Sampling}

As introduced in Section~\ref{sec:logretiever}, we proposed a hard negative sampling method to overcome the impact of overly easy negative examples. The Table~\ref{table:hardnega} presents the performance impact of hard negative sampling. The first row is the results of Log Retriever and the second row is results without hard negative sampling. We can see that hard negative sampling gives a 28\% performance improvement for HDFS top-$1$ setting and 24\% improvement for OpenSSH top-$1$ setting. 

The improvement in the Spark dataset is less than than the other two datasets. We can preserve 7\% of the accuracy gain through hard negative sampling.  A possible explanation for this might be that the Spark dataset is harder than the other two datasets, which leads to random negative samples that are also not overly simple. Therefore, the gain from hard negative sampling in Spark data would be less than HDFS and OpenSSH datasets.

\subsubsection{Hard Negative Example Weight} 
\begin{figure}[ht]
	\begin{center}
		\includegraphics[width=8.5cm]{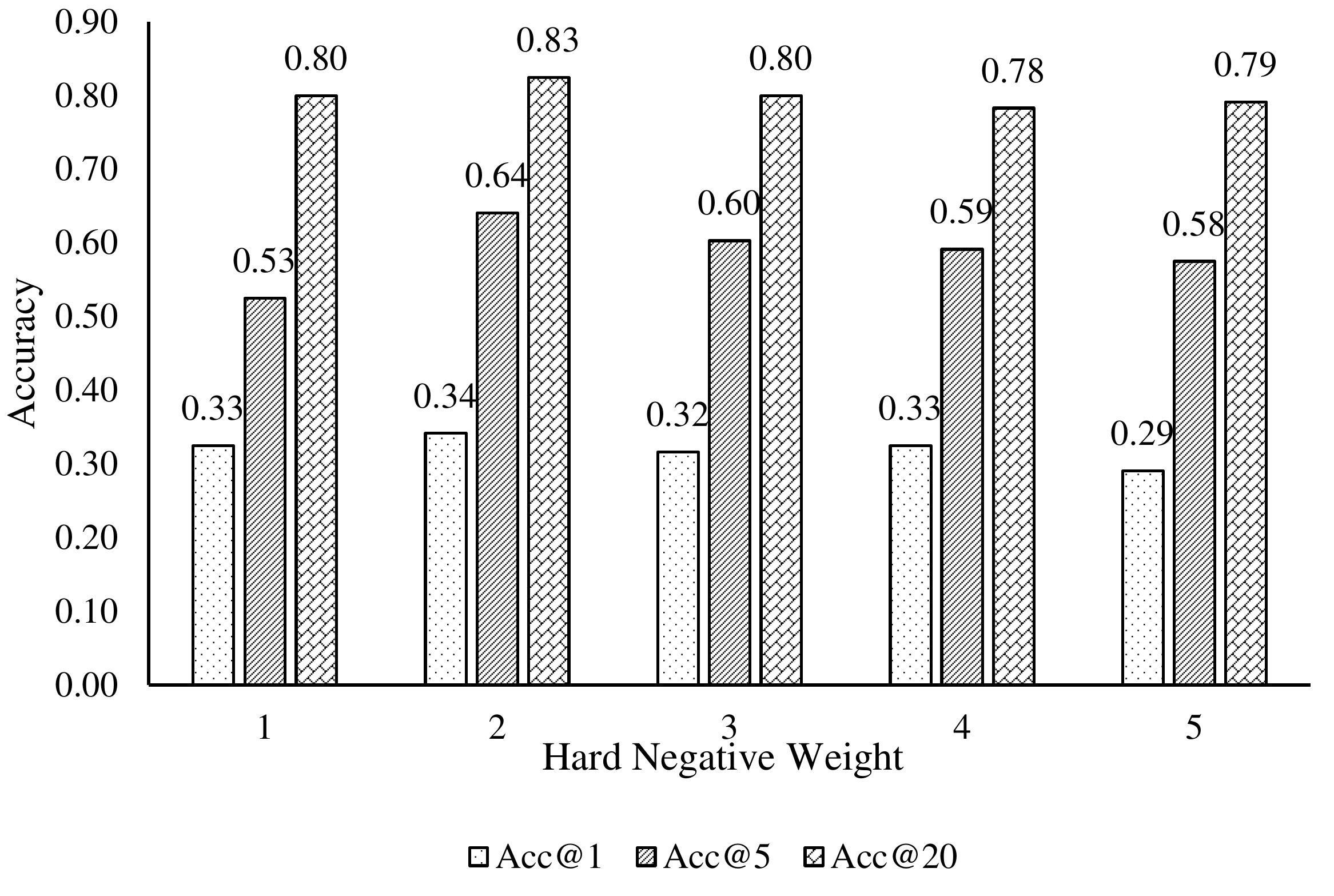}
	\end{center}
	\caption{Impact of different hard negative weights.}
	\label{fig:hn_weight}
\end{figure}

Based on Equation~\ref{equa:logretrieve_hard}, we incorporate the weighting of different negatives into our training. In this section, we investigate the performance impact of different hard negative example weights. We conduct this analysis experiment on the Spark dataset. The Figure~\ref{fig:hn_weight} shows that the Log Retriever model is not very sensitive to different hard negative example weights in the top $1$. Accuracy at top-$5$ varies widely according to different weights. The results demonstrate that we don't need to tune this hyper-parameter if we focus on Acc@1.

\begin{figure}[ht]
	\begin{center}
		\includegraphics[width=8.5cm]{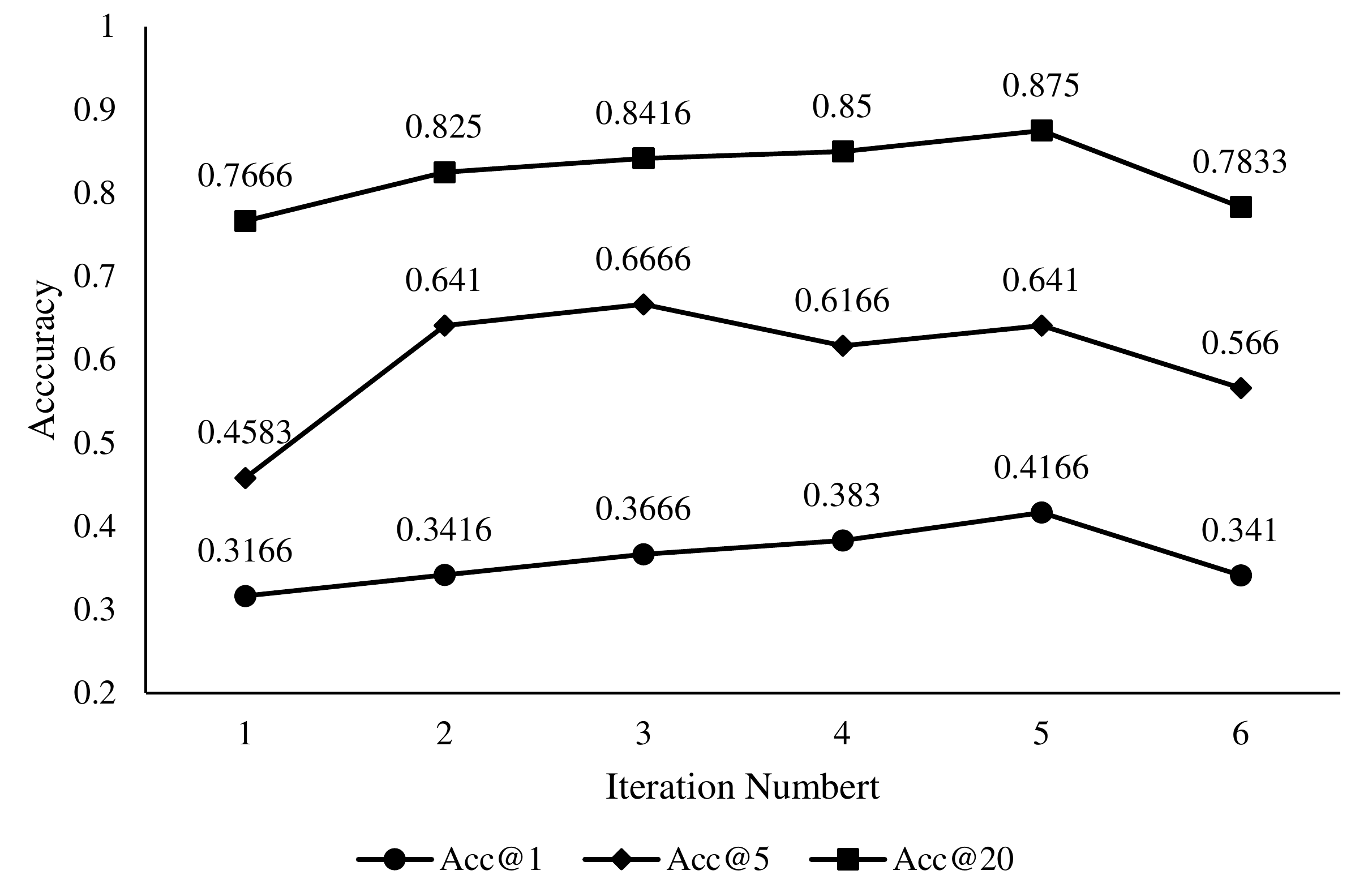}
	\end{center}
	\caption{Log retrieval accuracy with different iteration numbers.}
	\label{fig:iter_num}
\end{figure}

\begin{table*}[t]
\renewcommand{\arraystretch}{1.8}
\setlength{\tabcolsep}{12pt}
\centering
\caption{Evaluation results of log reading.
    }
    \centering
    \label{table:logreader}
\begin{tabular}{|l|ll|ll|ll|}
\hline
                       & \multicolumn{2}{|c|}{\textbf{HDFS}}                 & \multicolumn{2}{c|}{\textbf{OpenSSH}}              & \multicolumn{2}{c|}{\textbf{Spark}}                \\ \hline \hline
                       & \textbf{EM} & \textbf{F1} & \textbf{EM} & \textbf{F1} & \textbf{EM} & \textbf{F1} \\ \hline
\textbf{Random Guess}                  & 0.0933         & 0.1126         & 0.0689          &  0.0919         & 0.0416         &  0.0722          \\
\textbf{Slide Window}           & 0.0133         & 0.0281         & 0.0689          & 0.0747        & 0.0250         &  0.1348        \\
\textbf{Logistic Regression}    &0.2266 & 0.2322 & 0.1896 & 0.2011 & 0.1083 & 0.1510 \\
\textbf{BiDAF}                  &0.3866 & 0.4133 & 0.3275 & 0.3353 & 0.1666 & 0.2532 \\
\textbf{QANet}                  &0.4012 & 0.4248 & 0.3125 & 0.3291 & 0.2000 & 0.3121 \\
\textbf{BERT}                  &0.4362 & 0.4523 & 0.4154 & 0.4251 & 0.2333 & 0.3210 \\
\textbf{Log Reader}             &\textbf{0.4933} & \textbf{0.4933} & \textbf{0.4310} & \textbf{0.4484} & \textbf{0.3000} & \textbf{0.4486}          \\ \hline
\end{tabular}
\end{table*}

\begin{figure}[ht]
	\begin{center}
		\includegraphics[width=8cm]{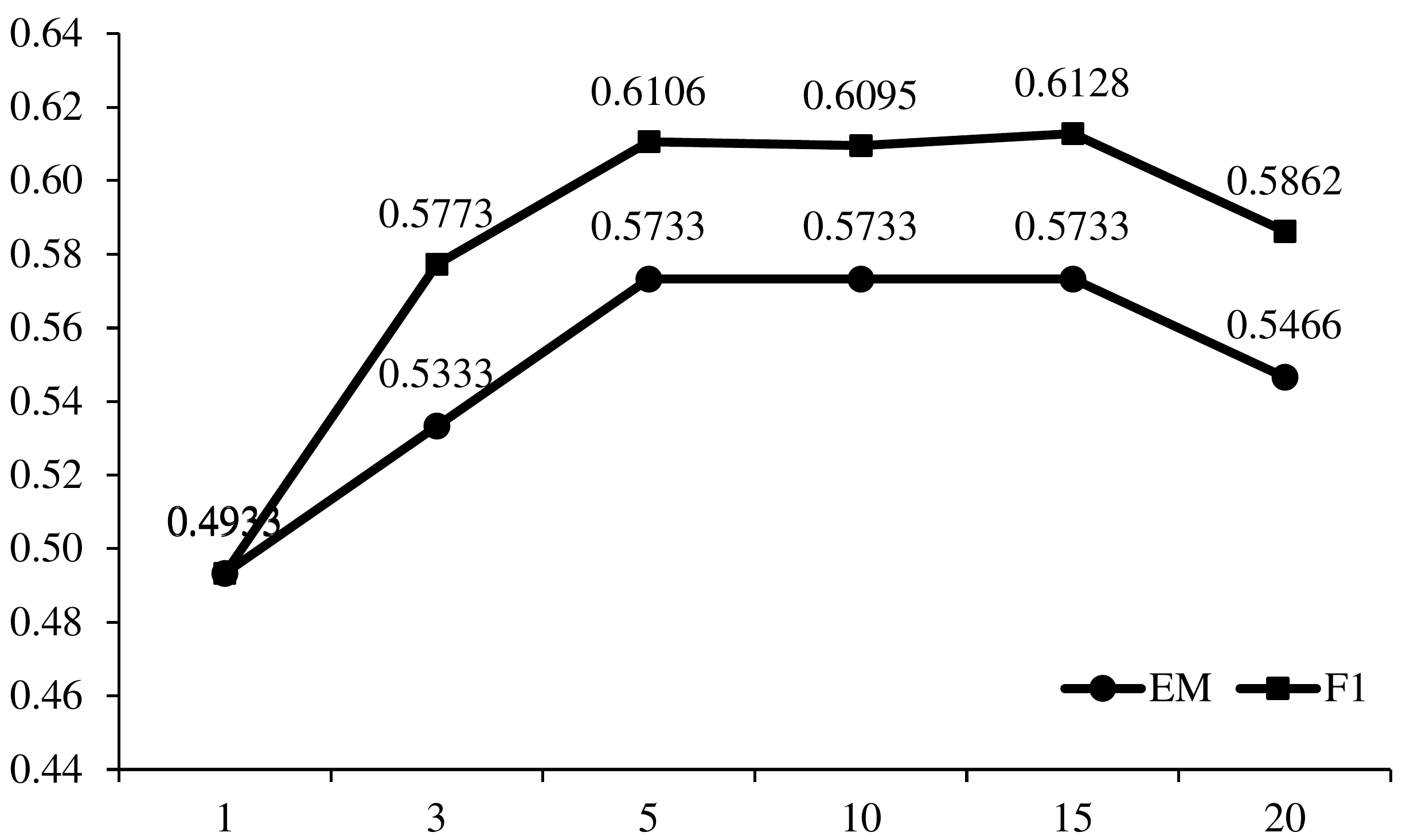}
	\end{center}
	\caption{Log Reader evaluation with different number of retrieved logs in HDFS dataset.}
	\label{fig:hdfs_layer}
\end{figure}

\begin{figure}[ht]
	\begin{center}
		\includegraphics[width=8cm]{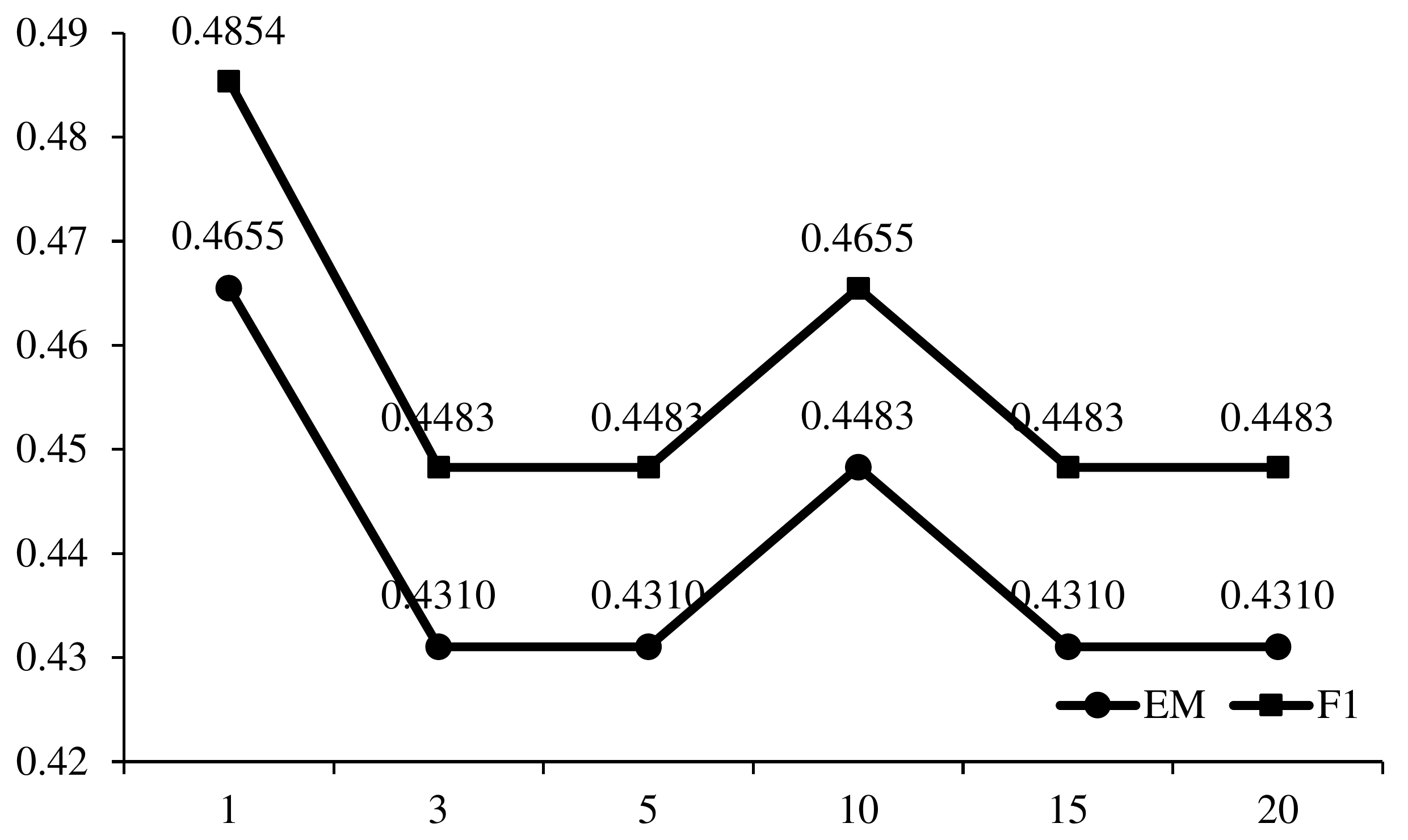}
	\end{center}
	\caption{Log Reader evaluation with different number of retrieved logs in OpenSSH dataset.}
	\label{fig:openssh_layer}
\end{figure}

\begin{figure}[ht]
	\begin{center}
		\includegraphics[width=8cm]{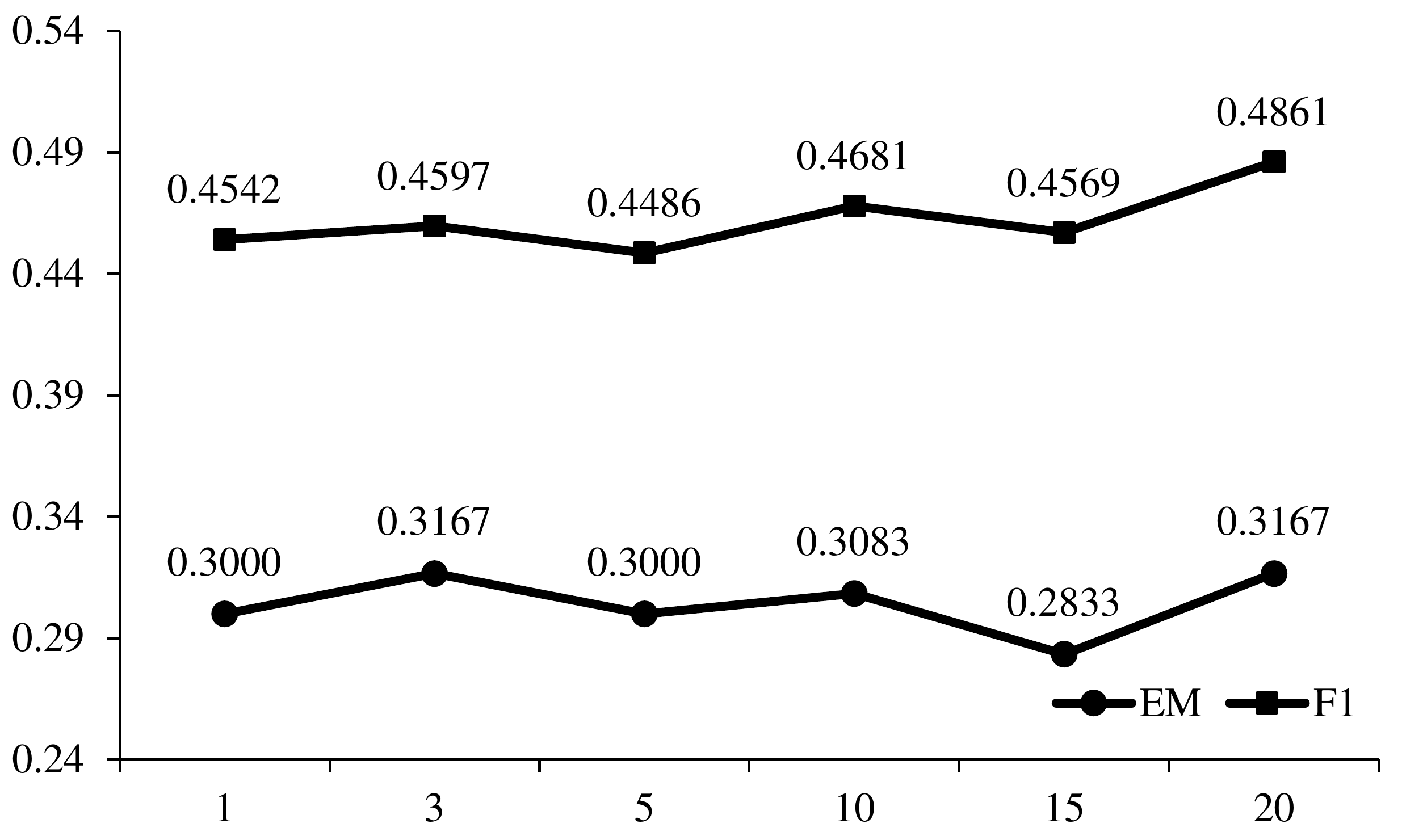}
	\end{center}
	\caption{Log Reader evaluation with different number of retrieved logs in Spark dataset.}
	\label{fig:spark_layer}
\end{figure}

    

\subsubsection{Different Iteration Numbers}

As shown in Figure~\ref{fig:iter_num}, we increase the training iteration number while keeping the default values for others and observe its results over the Spark dataset. We can see that the accuracy of the Log Retriever model increases with the training iteration at the beginning. However, after five epochs, the performance of log retrieving declines. There are several possible explanations for this result. First, the model may have been overfitted by the training dataset due to the limited size. Second, after several training epochs, sampled hard negatives are difficult for the model to learn, which has negative impacts on the performance.

\subsection{Log Reader}

In this part, we first study the performance of Log Reader on the three log datasets.  Then we present the experiments with a different number of retrieved logs. Lastly, we investigate the impacts of different log retrieval models.

\subsubsection{Performance of Log Reader} In our experiments, we use Log Retriever to retrieve log candidates from raw log corpus and feed top-5 logs into log reading models. Table~\ref{table:logreader} shows the performance of Log Reader compared to six baseline methods over three log datasets. In the log reading task, our proposed model achieves the highest EM and F1 scores over three datasets. On the HDFS dataset, the EM of Random Guess is low, which implies that the log reading task is challenging. We can see that Slide Window has even worse EM and F1 than Random Guess. A possible explanation is that answers often have no overlap with the question. On the OpenSSH dataset, Log Reader achieves the best performance among those methods. We observe that the EM and F1 of our model on OpenSSH are lower than the ones on the HDFS dataset. As shown in Table~\ref{table:logre}, the accuracy of Log Retriever on OpenSSH is also 6 points lower than the accuracy on HDFS. This experiment demonstrates that achieving high performance on OpenSSH is more challenging than HDFS. The data in Spark dataset is more diverse and contains more types of questions than the other two datasets. Our method achieves an EM score of 0.3 and an F1 of 0.4486  on Spark dataset, respectively.

\begin{figure}[ht]
	\begin{center}
		\includegraphics[width=8.5cm]{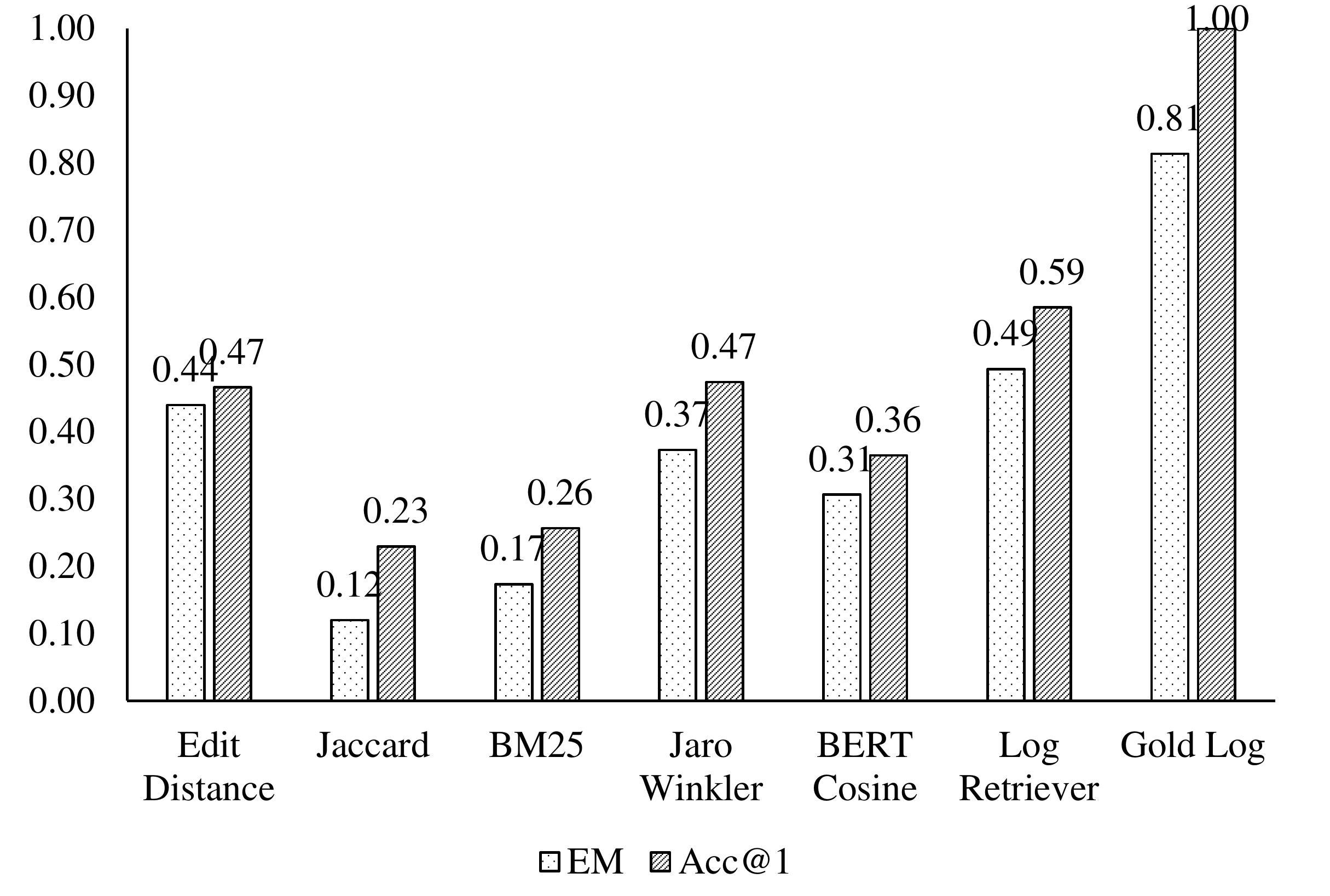}
	\end{center}
	\caption{Log reading accuracy with different log retriever models.}
	\label{fig:different_retriever}
\end{figure}

\subsubsection{Different number of retrieved logs}
In this experiment, we explore how many retrieved logs examples are needed to achieve good log reading performance. Figure~\ref{fig:hdfs_layer}-\ref{fig:spark_layer} illustrates the performance of Log Reader with respect to different numbers of retrieved logs examples on three datasets. There is a different pattern among these three datasets. As shown, on the HDFS dataset, Log Reader performs the best when using only 5, 10, or 15 examples. On the other side, Log Reader achieves the best performance when using top-1 retrieved log on the OpenSSH dataset. On the Spark dataset, Log Reader performs relatively stable and its best results occur in the top-20 setting. This suggests that a different number of retrieved logs  should be used in different datasets. The number of retrieved logs has a major impact on the performance of log reading tasks. A proper number of retrieved logs can improve the log reading accuracy.

\subsubsection{Different log retriever models} In this part, we present the performance with different log retriever models. Figure~\ref{fig:different_retriever} demonstrates our Log Reader's performance results on the HDFS dataset, measured by EM with the retrieved logs from different log retrieval models. The left bar is an exact match score of Log Reader and the right is accuracy at top-1 of log retrieval models. We can see that a higher retriever accuracy typically leads to better final QA results. Edit Distance and Jaro Winkle achieve the same accuracy scores of 0.47, but answers extracted from the logs retrieved by Edit Distance are more likely to be correct.

We don't use retriever to generate log candidates and directly use logs that contain the correct answers as gold logs. We feed gold logs into the Log Reader. We can see the impact of gold log in Figure~\ref{fig:different_retriever}.  Our experiments on HDFS data show that gold log achieves the highest EM score and it outperforms Log Retriever in reading by almost 30 points. There is still plenty of room to improve the log retrieval.

\subsection{Case Study}

\begin{table*}[t]
\renewcommand{\arraystretch}{1.5}
\setlength{\tabcolsep}{12pt}
\centering
\caption{A case study of log retrieval on HDFS and Spark logs.
    }
    \centering
    \label{table:logretriever_example}
\begin{tabular}{ll}
\hline
\textbf{HDFS}          &                                                                                                  \\ \hline
Question      & What is the block that is receiving from 10.251.123.132:57542?                                   \\
Edit Distance & Verification succeeded for blk\_1150231966878829887                                              \\
Jaccard       & Receiving block blk\_1744485040428751334 src: /10.250.5.237:52234 dest:   /10.250.5.237:50010    \\
BM25          & PacketResponder 2 for block blk\_8229193803249955061 terminating                                 \\
Jaro Winkler  & PacketResponder 1 for block blk\_2151150262081352617 terminating                                 \\
BERT Cosine   & Received block blk\_155320394753274773 of size 67108864 from   /10.251.201.204                   \\
Log Retriever & Receiving block blk\_-28342503914935090 src: /10.251.123.132:57542 dest:   /10.251.123.132:50010 \\ \hline \hline
\textbf{Spark}         &                                                                                                  \\ \hline
Question      & How many ms did it take to read the broadcast variable 37?                                       \\
Edit Distance & Started reading broadcast variable 37                                                            \\
Jaccard       & Reading broadcast variable 3 took 17 ms                                                          \\
BM25          & Got assigned task 3                                                                              \\
Jaro Winkler  & Block broadcast\_0 stored as values in memory (estimated size 384.0 B,   free 317.5 KB)          \\
BERT Cosine   & Block broadcast\_8\_piece0 stored as bytes in memory (estimated size 21.4   KB, free 35.4 KB)    \\
Log Retriever & Reading broadcast variable 37 took 14 ms          \\ \hline \hline                                     
\end{tabular}
\end{table*}

\begin{table*}[t]
\renewcommand{\arraystretch}{1.5}
\setlength{\tabcolsep}{12pt}
\centering
\caption{A case study of log reading on  Spark logs.
    }
    \centering
    \label{table:logreader_example}
\begin{tabular}{ll}
\hline
Question            & What is the estimated size of   the block broadcast\_27?                                       \\ \hline
Top-5 Logs          & Block broadcast\_25 stored as values in memory (estimated size   10.1 KB, free 419.6 KB)       \\
                    & Block broadcast\_27 stored as values in memory (estimated size   9.2 KB, free 404.2 KB)        \\
                    & Block broadcast\_26 stored as values in memory (estimated size   9.7 KB, free 389.6 KB)        \\
                    & Block broadcast\_27\_piece0 stored as bytes in memory (estimated   size 5.4 KB, free 395.0 KB) \\
                    & Block broadcast\_28\_piece0 stored as bytes in memory (estimated   size 5.6 KB, free 409.8 KB) \\ \hline
Logistic Regression & Block broadcast\_25 stored as values in memory (estimated size 10.1 KB, free \textcolor{red}{\textbf{419.6}} KB)       \\
BiDAF               & Block broadcast\_27\_piece0 stored as bytes in memory (estimated   size \textcolor{red}{\textbf{5.4}} KB, free 395.0 KB)  \\
QANet               & Block broadcast\_25 stored as values in memory (estimated size \textcolor{red}{\textbf{10.1}} KB, free 41.96 KB)  \\
BERT               & Block broadcast\_27 stored as values in memory (estimated size   \textcolor{red}{\textbf{9.2}} KB, free 404.2 KB)   \\
Log Reader          & Block broadcast\_27 stored as values in memory (estimated size   \textcolor{red}{\textbf{9.2}} KB, free 404.2 KB)   \\      \hline                                                                                 
\end{tabular}
\end{table*}

In the Table~\ref{table:logretriever_example}, we show some log retrieval examples on HDFS and Spark datasets. For the HDFS data, we can find that only Jaccard, BERT Cosine, and Log Retriever can rank a very related log candidate (logs about `\textit{receiving}' events) at the top-1 position. In this case, IP addresses and IP ports are domain-specific symbols, Jaccard and BERT Cosine cannot identify these symbols. Another case is from the Spark dataset as shown in Table~\ref{table:logretriever_example}. The question is \textit{``How many ms did it take to read the broadcast variable 37?''}, where two key factors are the action \textit{read} and the broadcast variable \textit{37}. The Jaccard method mixed broadcast variables up and other baseline methods selected a wrong log event.

We also present the examples on the log reading task in Table~\ref{table:logreader_example}. In this case, the question is \textit{What is the estimated size of the block broadcast\_27?} and we use top-5 logs retrieved from Log Retriever as our log candidates. We can find that top-5 logs are very similar and relevant to the question, but only the second log is the correct log candidate. We present the results of Logistic Regression, BiDAF, QANet, BERT and Log Reader, where the red color denotes the predicted answer. One thing worth noticing is that Log Retriever provides a wrong log at top-1 position, but BERT and Log Reader are able to select the correct answer from the correct logs. It is because the probability of the selected answer in the correct log is higher than in the top-1 log. It proves the robustness of our proposed methods.

\section{Related Works}\label{sec:related}
A large number of QA benchmarks have been released in the past decade. As introduced in~\cite{zhu2021retrieving}, the information source of the datasets are mainly from Wikipedia, Search Engines (such as Bing, Google, and Baidu), Online News (such as CNN/DailyMail), and Internet Forum (such as Reddit and Stack Overflow). Different from general corpora, unstructured logs belong to specific domains. A log message is composed of a message header and message content. The message header is designed by the logging framework and has a fixed structure.  In contrast, the message content is mainly written by developers in free-form natural language~\cite{he2017towards}, including the fixed text written by the developers (e.g., \textit{Received block * of size * from *}) and the values of the program variables to carry dynamic run-time information. Practically, modern systems produce large volumes of logs. Consequently, it is much more difficult to find the correct answers in unstructured logs.

For purposes other than ours, other types of log question answering systems have been proposed. Burkhardt~\cite{burkhardt2016quark} proposed an architecture and a prototypical first partial implementation to answer customer questions in the telecommunication domain from several knowledge sources. The author extracted a set of FAQs (Frequently Asked Question) from the telecommunication logs. Given a new question, the system searches for the closest question from FAQs. Due to the ever-increasing variety of logs, this system cannot answer unknown questions. Vanessa al.~\cite{lopez2007aqualog} proposed an approach to leverage ontology-based semantic markup to answer the questions in unstructured logs. It takes queries expressed in natural language and ontology as input and returns answers drawn from knowledge bases, which are contracted from unstructured logs. However, maintaining the knowledge bases becomes increasingly difficult as the system logs grow.
 
\section{Conclusion}\label{conc}
In this paper, we proposed LogQA, a log-based question answering system. We decomposed LogQA into two steps: retrieve, and then read. Given a question, Log Retriever search for potentially helpful logs from a large log corpus. We designed an interactive training method to involve hard negatives to improve the performance of the similarity model. Given the top $k$ retrieved logs, Log Reader extracts the answer span and selects the answer with the highest probability. Due to the lack of a public question answering dataset, we manually labeled the above three public log datasets and will make them publicly available. We evaluated our proposed system on these log datasets. The results demonstrated that our method outperforms other baseline log-based methods.

One of the future directions of our work is to construct more complex questions like those requiring multi-hop reasoning. For example, given a question, the retriever aims to search for the relevant logs from a large corpus in multiple steps.


%




\ifCLASSOPTIONcaptionsoff
  \newpage
\fi


\bibliography{tnsm_2020}
\bibliographystyle{IEEEtran}

\end{document}